\documentclass[a4paper,12pt]{article}
\usepackage{latexsym}

\begin{document}

\title{Every timelike geodesic in anti--de Sitter spacetime is a circle of the same radius}
\author{Leszek M. SOKO\L{}OWSKI and Zdzis\l{}aw A. GOLDA \\
Astronomical Observatory, Jagiellonian University,\\ 
Orla 171,  Krak\'ow 30-244, Poland\\
and Copernicus Center for Interdisciplinary Studies,\\ 
email: lech.sokolowski@uj.edu.pl,\\
email: zdzislaw.golda@uj.edu.pl} 

\date{}
\maketitle

\begin{abstract}
We refine and analytically prove an old proposition due to Calabi and Markus on the shape 
of timelike geodesics of anti--de Sitter space in the ambient flat space. We prove that 
each timelike geodesic forms in the ambient space a circle of the radius determined by 
$\Lambda$, lying on a Euclidean two--plane. Then we outline an alternative proof for 
$\textrm{AdS}_4$. We also make a comment on the shape of timelike geodesics in de Sitter 
space.
Pacs: 04.20Jb 
\end{abstract}

1. In 1962 Calabi and Markus wrote in a famous paper \cite{CM}:\\ 
`Each geodesic on $\textrm{AdS}_n$ is a component of the intersection of $\textrm{AdS}_n$ with a 
2--plane through the origin in the ambient $\mathbf{R}^{n+1}$. Every 2--plane in 
$\mathbf{R}^{n+1}$ through the origin O, meeting $\textrm{AdS}_n$, inherits a metric which is 
either negative definite, indefinite, or negative semi--definite of rank 1, since it contains 
a negative definite line. According to the three cases in their respective order, the 
corresponding geodesics are timelike (ellipses), space--like (branch of hyperbola), or 
light--like (straight lines).'\\ 
The statement, contained in the introductory part of the paper, was left without a detailed 
proof since the paper was focussed on de Sitter space. The statement was later repeated 
several times in the literature and merely graphically illustrated without any analytic 
proof (see e.~g.~\cite{Mo}). We have encountered claims that the theorem immediately 
follows from the geometrical structure of $\textrm{AdS}_d$ for $d\geq 4$ and of the embedding 
space. Yet other (more numerous) relativists claim that it, though known, is not quite obvious 
and in particular the fact that all timelike geodesics are a circle (and not general 
ellipses) of the same radius, deserves an explicit analytic proof.\\

2. Here we show that each timelike geodesic in $d$--dimensional anti--de Sitter space 
($d=n+1$ with $n\geq 3$)  
is a circle of radius $a=[-\frac{1}{2}n(n-1)/\Lambda]^{1/2}$ lying on a Euclidean two--plane 
in the embedding space, rather than the ellipse proper; $\Lambda<0$ is the cosmological 
constant. We consider the proper anti--de Sitter space $\textrm{AdS}_d$ (and not its 
covering space) defined as a pseudosphere (a quadric)
\begin{equation}\label{n1}
\eta_{AB}\,X^A X^B= -\sum^n_{i=1} (X^i)^2+U^2+V^2=a^2
\end{equation}
in the ambient flat space $\mathbf{R}^{n,2}$ with Cartesian coordinates $X^A=(X^i,U,V)$, 
$A,B=1,\ldots,n+2$, endowed with the ultrahyperbolic metric  
\begin{equation}\label{n2}
ds^2=\eta_{AB}\,dX^A dX^B= -\sum^n_{i=1} (dX^i)^2 +dU^2+dV^2,
\end{equation}
where $\eta_{AB}=\textrm{diag}[-1,\ldots,-1,1,1]$. Any timelike geodesic on $\textrm{AdS}_d$ 
is a curve G in $\mathbf{R}^{n,2}$ with a timelike tangent unit $(n+2)$--vector 
$u^A\equiv \dot{X}^A=dX^A/ds$, $\eta_{AB}\,u^A\,u^B=1$. For any geodesic the standard 
Lagrangian $\sqrt{\eta_{AB}u^Au^B}$ is equivalent to $\frac{1}{2}\eta_{AB}u^Au^B$. In order 
to have a timelike geodesic on $\textrm{AdS}_d$ described as a curve G in the ambient 
space, one applies the Lagrange multipliers method and the Lagrangian reads
\begin{equation}\label{n3}
L(X^A,\dot{X}^A,\lambda)=\frac{1}{2}\eta_{AB}\,\dot{X}^A \dot{X}^B+\lambda(s) F(X^A),
\end{equation}
where the constraint is 
\begin{equation}\label{n4}
F(X^A)=\eta_{AB}\,X^A X^B-a^2=0. 
\end{equation}
The Lagrange equations are then $F=0$ and 
\begin{equation}\label{n5}
\ddot{X}^A-2\lambda X^A=0.
\end{equation}
To eliminate the multiplier $\lambda$ from (5) one takes the second derivative of the 
constraint $F=0$ what yields 
\begin{equation}\label{n6}
\eta_{AB}\,\ddot{X}^A X^B= -1.
\end{equation}
Next multiplying eq. (5) by $\eta_{AB}X^B$ and applying (6) and (1) one arrives at 
$\lambda=-(2a^2)^{-1}$ and the Lagrange equations reduce to 
\begin{equation}\label{n7}
\ddot{X}^A+\frac{1}{a^2} X^A=0.
\end{equation}
The general solution is 
\begin{equation}\label{n8}
X^A=q^A\sin \frac{s}{a}+r^A\cos\frac{s}{a},
\end{equation}
where the constant vectors $q^A$ and $r^A$ are subject to $q^Ar_A=0$ and $q^A q_A=
r^A r_A=a^2$. \\
Now the proof that the curve is a circle on a plane is immediate if one employs the full 
$SO(n,2)$ symmetry of the $\mathbf{R}^{n,2}$. Let $P_0\in \mathbf{R}^{n,2}$ 
be an initial point ($s=0$) of an arbitrary timelike geodesic G on the $\textrm{AdS}_d$, 
the geodesic is future and past extended from $P_0$. Take any transformation of $SO(n,2)$ 
which makes the coordinates of  $P_0$ equal $X^i(P_0)=0=U(P_0)$ and $V(P_0)=a$, the 
transformation is non--unique. Then by the remaining transformations leaving invariant 
the straight line joining $P_0$ with the origin $X^A=0$ one makes the tangent to G 
at $P_0$ vector $\dot{X}^A(0)$ tangent to the  $U$ line through $P_0$, 
i.~e.~$\dot{X}^i(0)=0=\dot{V}(0)$ and $\dot{U}(0)=1$. Then the parametric 
representation of G is reduced to 
\begin{equation}\label{n9}
X^i(s)=0, \qquad U(s)=a\sin\frac{s}{a}, \qquad V(s)=a\cos\frac{s}{a}.
\end{equation}
Each timelike geodesic on $\textrm{AdS}_d$ is represented in the ambient $\mathbf{R}^{n,2}$ 
by a circle of radius $a$ on an appropriately chosen Euclidean two--plane $(U,V)$. (To 
avoid any doubts we stress that a circle is defined as the locus of points at a fixed 
proper distance away from a reference point, in this case $X^A=0$; this definition is 
invariant under the symmetry transformations of $\mathbf{R}^{n,2}$.) In general (the 
coordinate system $X^A$ is not adapted to the initial conditions of G) each timelike 
geodesic of AdS space is the circle lying on a Euclidean two--plane going through the 
origin $X^A=0$ of the ambient space. In general two timelike geodesics do not intersect 
and this means that their two--planes do not intersect either and the planes have only 
one common point, the origin. The circle is closed 
in full conformity with the well known fact that a time coordinate on the proper 
$\textrm{AdS}_d$ is periodic and each timelike geodesic has the length $2\pi a$ in the 
period.\\
The theorem shows that the standard distinction between radial, circular and `general' 
timelike geodesics (commonly made in static spherically symmetric spacetimes) in the case 
of anti--de Sitter space has no deeper geometrical meaning and is merely coordinate 
dependent (in AdS). In this spacetime there is only one kind of timelike geodesics with 
their curvature determined by the spacetime curvature.\\

3. It is interesting to notice that there is an alternative proof of the theorem which 
makes no use of the $SO(n,2)$ symmetry of $\mathbf{R}^{n,2}$, but it is laborious. We 
sketch it in the case of $\textrm{AdS}_4$. 
To determine the shape of a generic timelike geodesic G in $\mathbf{R}^{3,2}$ we employ 
the Frenet--Serret formalism \cite{Sp,IV} and construct a pentad $\{e_K\}$, $K=1,\ldots,5$, 
of orthonormal five--vectors adapted to G. To this end one must express the Cartesian 
components $e_K{}^A$ of the vectors in terms of a coordinate system on $\textrm{AdS}_4$ 
and the latter are functions of some parameter along G. The coordinates 
$X^A=(X,Y,Z,U,V)$ of $\mathbf{R}^{3,2}$ on $\textrm{AdS}_4$ are most conveniently 
parametrized  by the static coordinates $x^{\alpha}=(t,r,\theta, \phi)$ as 
	\begin{eqnarray}\label{n10}
X & = & r\sin\theta\cos\phi,\quad Y=r\sin\theta \sin\phi, \quad Z=r\cos\theta,
\nonumber\\
U & = & \sqrt{r^2+a^2}\,\sin\frac{t}{a}, \quad V=\sqrt{r^2+a^2}\,\cos\frac{t}{a}, 
	\end{eqnarray} 
here $r\in[0,\infty)$, $t/a\in(-\pi,\pi]$ and $t$ is a time coordinate on a circle giving 
rise to closed timelike lines. In practice these coordinates cover the entire  
manifold and its metric reads
\begin{equation}\label{n11}
 ds^2=\frac{r^2+a^2}{a^2}dt^2-\frac{a^2}{r^2+a^2}\,dr^2-r^2\,(d\theta^2+
\sin^2\theta\,d\phi^2). 	
\end{equation}
(To be strict we admit that the static chart does 
not cover the whole manifold in the proper sense since the variables are defined on 
half-open sets and there are the standard singularities of spherical coordinates.)
The chart is fixed and only rotations of the $S^2$ spheres are admissible, 
then any timelike geodesic G has $\theta(s)=\pi/2$. 
The proper time 
derivatives $\dot{t}$ and $\dot{\phi}$ are eliminated with the aid of the integral of 
energy per unit mass of the particle travelling on the geodesic, $k$, and the conserved 
angular momentum per unit mass, $h$,
	\begin{equation}\label{n12}
 \dot{t}=\frac{a^2k}{r^2+a^2}, \qquad  \dot{\phi}=\frac{ah}{r^2}.
	\end{equation} 
$k>0$ and $h\geq 0$ are dimensionless and subject to $M\equiv k^2-1-h^2>0$. 
Excluding the almost trivial case of circular geodesics, $r=\textrm{const}$, for 
which the procedure below gives the same final outcome, we assume that each G is 
parametrized by the radial coordinate, $x^{\alpha}=x^{\alpha}(r)$. The geodesic 
is confined to the range $r_{\rm min}\leq r\leq r_{\rm max}$ where 
$r^2_{\rm min, max}=\frac{1}{2}a^2(M\mp\sqrt{M^2-4h^2})$. (If the geodesic is 
located at the singularity $r=0$, then $h=0$ and $k=1$.) In other terms, in the fixed 
frame one has a set of timelike geodesics parameterized by $k$ and $h$.\\

The first vector of the pentad is identified with the tangent one to G, 
 $e_1{}^A\equiv u^A=dX^A/ds$, then 
	\begin{eqnarray}\label{n13}
e_1{}^A&=&\left[\dot{r}\cos\phi-\frac{a\,h}{r}\sin\phi,\dot{r}\sin\phi+\frac{a\,h}{r}\cos\phi,0,\right.\nonumber\\
&&\left.
\frac{1}{\sqrt{r^2+a^2}}\left(
r\dot{r}\sin\frac{t}{a}+a\,k\cos\frac{t}{a}
\right),
\frac{1}{\sqrt{r^2+a^2}}\left(
r\dot{r}\cos\frac{t}{a}-a\,k\sin\frac{t}{a}
\right)
\right],\nonumber\\
	\end{eqnarray} 
where $\dot{r}\neq 0$ is determined from the integral of motion $g_{\alpha\beta}\dot{x}^{\alpha}
\dot{x}^{\beta}=1$. Clearly $e_1\cdot e_1\equiv \eta_{AB}\,e_1{}^Ae_1{}^B=1$. Then $\dot{e}_1$ 
is orthogonal to $e_1$ and determines the second vector, $e_2$, by $\dot{e}_1\equiv\kappa 
e_2$, where $\kappa>0$ is the curvature of G. One finds 
\begin{equation}\label{n14}
 e_2{}^A=-\frac{1}{a}\left[r\cos\phi, r\sin\phi, 0, (r^2+a^2)^{1/2}\sin\frac{t}{a}, 
 (r^2+a^2)^{1/2}\cos\frac{t}{a}\right],
\end{equation} 
with $e_2\cdot e_2=+1$ and $\kappa=1/a$. The local 2--plane $\Pi(s)$ spanned on $e_1(s)$ 
and $e_2(s)$ is Euclidean. The vector $e_3$ is undetermined by the formalism since 
$\dot{e}_2=-\frac{1}{a}e_1$ and the first, 
second and third torsion of the curve vanish, $\tau_1=\tau_2=\tau_3=0$. At this 
point to complete the proof it is sufficient to apply two theorems:\\ 
i) if the first torsion is identically zero (and in consequence also 
$\tau_2=\tau_3=0$ everywhere) in a flat ambient five--space, then the curve lies in a 
fixed 2--plane in that space (in other words the local plane $\Pi(s)$ is actually a 
constant plane $\Pi$ spanned on the variable vectors $e_1$ and $e_2$) \cite{Sp};\\ 
ii) a flat curve is uniquely determined (up to an isometry of the plane containing it) 
by its curvature and if the plane is Euclidean and $\kappa=\textrm{const}>0$, then 
the curve is an arc of a circle of radius $1/\kappa$.\\

On the other hand to find out the explicit equation of G on the plane requires an 
astonishingly complicated calculation. 
To this end we first determine the remaining vectors $e_3$, $e_4$ and $e_5$ of the pentad 
by applying the orthonormality conditions; these are normalized to $-1$. Obviously 
$e_3{}^A=[0,0,1,0,0]$ whereas $e_4$ and $e_5$ are so complex that can be found only with 
the aid of the program Mathematica; we shall not display them here. The vectors are constant 
ensuring that the planes $\Pi(s)$ coincide for all $s$ forming one Euclidean plane $\Pi$.  
The plane is tilted in the coordinates $X^A$ and the geometrical interpretation of G is 
difficult. We are therefore forced to make use of the $SO(3,2)$ symmetry of 
$\mathbf{R}^{3,2}$ to rotate the coordinate 
system $X^A$ in such a way as to make $\Pi$ lie on the $(UV)$ plane. This is achieved by 
transforming $e_4$ and $e_5$ to the form $e_4{}^A=[0,1,0,0,0]$ and $e_5{}^A=[1,0,0,0,0]$. 
First we transform $e_5$, it may be chosen with $e_5{}^3=e_5{}^5=0$, i.~e.~it lies in the 
subspace $\mathbf{R}^{2,1}=\{(X,Y,U)\}$ being the Minkowski 3--space. The required form of 
$e_5$ is attained by first performing a spatial rotation $R_5$ and then a Lorentz boost 
$B_5$, together 
	\begin{eqnarray}\label{n15}
B_5R_5=\left(\!
	\begin{array}{ccc}
\displaystyle	\frac{P}{\sqrt{M}}&\displaystyle	0&\displaystyle-\sqrt{\frac{2}{M}}h\\[3ex]
	0			&	1&0\\[1ex]
\displaystyle	-\sqrt{\frac{2}{M}}h&\displaystyle0&\displaystyle\frac{P}{\sqrt{M}}
	\end{array}
	\!\right)
\left(\!
	\begin{array}{ccc}
\displaystyle	\frac{H_{-}}{P}&\displaystyle\frac{H_+}{P} &0\\[2ex]
\displaystyle	-\frac{H_+}{P}&\displaystyle\frac{H_-}{P} &0\\[2ex]
\displaystyle		0			&\displaystyle	0&\displaystyle1
	\end{array}
	\!\right),
	\end{eqnarray}
where the auxiliary functions of $k$ and $h$ are $M=k^2-1-h^2$, $P=\sqrt{k^2+h^2-1}$, 
$H_{\pm}=F/2\pm kh/F$ and 
	\begin{equation}\label{n16}
F(h,k)=(M+2h^2-\sqrt{M^2-4h^2})^{1/2}.
	\end{equation}  
Yet $e_4$ belongs to the subspace $\mathbf{R}^{2,2}=\{(X,Y,U,V)\}$ and the transformation 
$B_5R_5$ (elevated to the subspace) simplifies it a bit. Then one rotates the resulting $e_4$ 
in the plane $(UV)$ by $R_4$,
	\begin{eqnarray}\label{n17}
R_4  = \left(\!
	\begin{array}{cc}
	N\sqrt{M}P&-N\sqrt{M^2-4h^2} \\
	N\sqrt{M^2-4h^2}&N\sqrt{M}P 
	\end{array}
	\!\right),
	\end{eqnarray}
$N(h,k)=(2k^4-2h^2k^2-4k^2-2h^2+2)^{-1/2}$. Finally $e_4$ is subject to a Lorentz boost in 
Minkowski plane $(YV)$,
	\begin{eqnarray}\label{18}
R_4 & =  \left(\!\!
	\begin{array}{cc}
	\cosh\psi&-\sinh\psi \\
	-\sinh\psi&\cosh\psi 
	\end{array}
	\!\!\right),
	\end{eqnarray} 
with $\sinh\psi=-(\sqrt{2}NP)^{-1}$. Altogether the required transformation in $\mathbf{R}^{3,2}$ 
is $L=B_4R_4B_5R_5$ (the matrices are of rank 5). Under the action of $L$ the curve G is 
described in the transformed coordinates $X^A$ as $X=Y=Z=0$ and $U$ and $V$ are extremely 
complicated irrational functions of $s$, $k$ and $h$. A long computation by Mathematica 
assisted by hand manipulations allowed us to show that $U^2+V^2=a^2$ for all $s$; we presume 
that this is the most intricate parametric representation of the circle ever known. 
At the end we notice that the origin O of the static coordinates $x^{\alpha}$ ($t=0$, $r=0$) 
is replaced under $L$ by a new origin O' which is translated in space and forward in time with 
respect to O, in accordance with the fact that in $\textrm{AdS}_4$ there does not exist a 
translational Killing vector tangent to the spaces $t=\textrm{const}$.\\

4. \textbf{Timelike geodesics in de Sitter space}. We compare the above result with 
properties of timelike geodesics in de Sitter space; for simplicity we only consider 
dimension four. We use the Gaussian normal geodesic (GNG) coordinates  covering the 
whole manifold (`the whole' --- as above) and making the metric time dependent, 
\begin{equation}\label{n19}
 ds^2= d\tau^2-
\frac{1}{H^2}\cosh^2H\tau\,(d\chi^2+\sin^2\chi\,d\Omega^2). 	
\end{equation}
This chart is essential since the frequently used static coordinates $(t,r,\theta,\phi)$ cover 
only a half of the manifold deceptively suggesting that circular timelike geodesics are excluded 
\cite{SG}. Actually they do exist for the angular radial coordinate $\chi=\pi/2$ (and $\theta=\pi/2$) 
and their set is parametrized by the angular momentum $h\neq 0$.\\
De Sitter space is a pseudosphere in the Minkowski five--space ${\cal M}_5$ with the metric 
$\eta_{AB}=\textrm{diag}[1,-1,-1,-1,-1]$, $A=0,\ldots, 4$ and this fact allows one to determine 
invariant geometric features of timelike geodesics. Let Z be any timelike geodesic. By a $SO(1,4)$ 
transformation one makes the time $X^0$ coordinate line in ${\cal M}_5$ tangent to Z at an initial 
point $P_0$. Then by applying the standard parametrization of coordinates $X^A$ on $\textrm{dS}_4$ 
in terms of the GNG coordinates (see e.~g.~eq.~(2.9) in \cite{Sa}) one proves that the unique 
solution of the geodesic equation is $\tau=s$, $\chi=\pi/2=\theta$, $\phi=0$, that is, Z coincides 
with the time coordinate line in the appropriately chosen GNG chart. In ${\cal M}_5$ it corresponds 
to $X^0=H^{-1}\sinh Hs$, $X^3=H^{-1}\cosh Hs$, $X^1=X^2=X^4=0$. 
The geodesic lies on Minkowski plane $(X^0X^3)$ and forms a pseudocircle of radius $1/H$ (one 
branch of a `hyperbola'); using the F--S formulae one finds that its curvature is $\kappa=H$. 
We conclude that also in $\textrm{dS}_4$ space all timelike geodesics are of one kind, a 
branch of hyperbola in the ambient ${\cal M}_5$, as is stated in the introduction to \cite{CM}; 
the analytic proof may be easily extended to any dimension above four.\\

In conclusion, we have analytically proved and refined the already known fact that in maximally 
symmetric spacetimes there is only one kind of timelike geodesics with the curvature determined 
by the spacetime curvature. The distinction between radial, circular and `general' curves is 
merely coordinate dependent and may be cancelled by an appropriate change of the reference 
frame; they may always coincide with some time lines.\\

 \textbf{Acknowledgements} We are indebted to Barbara Opozda for helpful comments on Riemannian 
geometry.\\

\end{document}